\documentclass[%
 reprint,amsmath,amssymb,aps,pre,numerical,nofootinbib
]{revtex4-1}

\usepackage{graphicx}% Include figure files
\usepackage{color}% Inlcude figures with color
\usepackage{dcolumn}% Align table columns on decimal point
\usepackage{epstopdf}% Making epslatex possible
\usepackage{bm}% bold math
\usepackage[utf8]{inputenc}

\usepackage{xcolor}

\begin{document}

\preprint{APS/123-QED}

\title{Diffusion with Resetting Inside a Circle}% Force line breaks with \\
\author{Abhinava Chatterjee\textsuperscript{1,2},
Christos Christou\textsuperscript{1} and Andreas
Schadschneider\textsuperscript{1}}
\affiliation{\textsuperscript{1}Institute for Theoretical Physics,
University of Cologne, Z\"ulpicher Stra{\ss}e 77, D-50937 K\"oln,
Germany} \affiliation{\textsuperscript{2}Department of Physical Sciences, 
Indian Institute of Science Education and Research Kolkata, Mohanpur - 741246, India}

\date{\today}

\begin{abstract}
We study the Brownian motion of a particle in a bounded circular 2-dimensional domain, in search for a stationary target on the boundary of the domain. The process switches between two modes: one where it performs a two-dimensional diffusion inside the circle and one where it travels along the one-dimensional boundary. During the diffusion, the Brownian particle resets to its initial position with a constant rate $r$. The Fokker-Planck formalism allows us to calculate the mean time to absorption (MTA) as well as the optimal resetting rate for which the MTA is minimized. From the derived analytical results the parameter regions where resetting reduces the search time can be specified. We also provide a numerical method for the verification of our results.   
  
\end{abstract}

\maketitle

\section{Introduction}

A large range of phenomena are characterized by the  properties of first passage time of different stochastic processes \cite{Redner01}. A special case of them are related to search processes that take place in confinement and have long been an object of scientific research \cite{Querin}. They find applications among a large range of distinct fields such as microbiological systems, network theory and computer science \cite{Redner14,Gelenbe,Montanari}. In the last few years, the mean time to absorption (MTA) for many target processes inside different bounded domains have been evaluated. For most stochastic processes, an analytical expression in an arbitrary domain is rather difficult to establish \cite{Benichou05}. Nevertheless progress can be made by the study of general concepts like the reduction of dimensionality \cite{Holyst,Benichou09,Benichou07,Evans12dec} in well defined domains. Here we consider such a simple confining geometry inside of which a stochastic search process combining two-dimensional diffusion inside a circle with one-dimensional diffusion on the periodic boundary takes place. We ask ourselves how the implementation of a resetting mechanism, which forces the diffusive searcher to return to its initial position, will influence the MTA if we assume that the target on the boundary is perfectly absorbing.

A resetting mechanism like the one introduced in the previous paragraph may seem purely artificial but it is strongly inspired by patterns often encountered in nature. For example animals may relocate to previously visited positions in order to improve their searching strategy \cite{Boyer}. Stochastic processes performing resetting as well as effectively reducing their dimensionality find apllications in various systems. For macrobiological processes like foraging \cite{catalan} such a combination arises naturally, for instance, hunting along the shore of a lake and returning to the roost. At the same time reduction of dimensionality can be considered as fulfillment of a constraint in terms of a randomized search algorithm which in general profit from a restart mechanism \cite{Pal17}.

Our study builds on an increasing amount of works discussing stochastic processes with resetting \cite{Shkilev,Kusmierz2,Sharma} which became only recently a subject of study in the physics literature \cite{Zanette,Montero13}. The interest on this field was strongly boosted by the work of Evans and Majumdar who first analyzed the case of a one-dimensional Brownian motion modified by a resetting mechanism \cite{Evans11}. The search properties of this process have been studied by considering an absorbing target at the origin and calculating the \textit{mean first passage time} $T$ to the origin as a function of the \textit{resetting rate} $r$. It is clear that $T$ diverges as $T\sim r^{-1/2}$ for $r\rightarrow 0$, which recovers the well-known result that the mean first passage time for a purely diffusive particle is infinite. Also as $r\rightarrow\infty$, the particle hardly gets an opportunity to diffuse away from its starting position and remains localized for an indefinite time. As a result, $T$ diverges. It is evident that a minimum exists for $T$ since $T$ diverges in the two limits $r\rightarrow 0$ and $r\rightarrow \infty$.

This setting has shown the drastic effect of a resetting mechanism on the distribution of search times and has lead to several questions with regard to the dynamics of similar processes. So the generalization of this problem for the special cases of space-depending rates, resetting to a random position with a given distribution and to a spatial distribution of the target were presented in \cite{Evans112}. It has also been shown that the presented formalism can easily be expanded for the higher dimensional cases \cite{Evans14}. At the same time the properties of the non-equilibrium steady state have been evaluated for the special case of a diffusion inside a potential landscape \cite{APal}.  
 
Next to the archetype of diffusion processes with resetting several other stochastic processes have been analyzed in a resetting context. As an example, for the first order transition observed for a Levy flight process with resetting, the mean first passage time and the optimal search time have been evaluated in \cite{Kusmierz}. This very interesting property of phase transitions was treated extensive in \cite{Campos,Touchette} while even the thermodynamics of this process could be analyzed \cite{Fuchs}. Furthermore resetting is not only relevant for stochastic search processes as the work of Gupta et. al. \cite{Gupta,Gupta2} shows, where the effect of a persistent resetting mechanism on the time evolution of fluctuating surfaces has been studied.

In the recent past the important influence of memory in the foraging behavior of
macrobiological organisms was analyzed by Boyer et al.~in
\cite{Boyer17} where the properties of a stochastic process
resetting to a previously visited site has been evaluated. Next to
the evaluation of the resetting rate in the last years the optimal
resetting time distribution could also been determined \cite{Apal2}.
Lastly we have to note that in most of the presented models
the resetting mechanism could be considered as an \emph{external} mechanism. A variation of this property was presented in the work
of Falcao and Evans \cite{Falcao}, where the possibility of a
resetting relies on the \emph{internal} dynamics.

In the following we imagine that we have a stochastic searcher
starting from a position inside of a circle with radius $R$. At the
same time we have a stationary target on the boundary of the circle.
The stochastic searcher performs a two-dimensional Brownian motion
with the diffusion constant $D_2$ until it arrives at the boundary
of the circle. The searcher then sticks to the boundary and
undergoes a one-dimensional diffusion with diffusion constant $D_1$
along the boundary. With a constant rate $r$ the particle then
resets to the initial position inside the circle.
Then again the particle performs a two-dimensional diffusion until
it arrives at the boundary and so on. The process terminates when
the searcher finds the target. We assume here that the target has no
dimension which ensures that such a termination is possible only
during the one-dimensional diffusion phase. This assumption allows
to calculate the MTA exactly by using the
results derived in a large number of works on stochastic processes
with resetting in the last few years.

This paper is structured as follows: we will first remind the
readers about the properties of diffusion processes with stochastic
resetting. In Sec.~\textbf{II} we will hereto consider the case where
the initial position of the particle is on the boundary of the
circle and analyze the corresponding MTA. In
Sec.~\textbf{III} we focus on the case with an initial position at
the center of the circle and compare our results to previous works
in an intermittent setting \cite{Moreau11}. In Sec.~\textbf{IV} we analyze the properties for
general initial conditions. We will first derive an expression for
the gain potential. This provides the master
equation of the process which can then be solved in order to give us
the desired MTA. Furthermore, we will
characterize the different areas for which a resetting proves
beneficial, depending on the parameters $D_2$, $D_1$ and $r$. In
Sec.~\textbf{V} we will follow the same approach as in
Sec.~\textbf{IV} in order to analyze the properties of the
generalized process where resetting may even occur during the
two-dimensional diffusion. In the last section we will focus on open
questions and possible extensions of the present work.

%%%%%%%%%%%%%%%%%%%%%%%%%%%%%%%%%%%%%%%%%%%%%%%%%%%%%%%%%%%%%%%%%%%%%%%%%%%%%%%%%

\section{Diffusion with Resetting in one-dimensional periodic domain}

We start with the simpler problem of an initial position on the
boundary of the circle. In this case we do not need to consider the
two-dimensional diffusion as this case boils down to just a
one-dimensional diffusion process with resetting in a periodic
domain of length $L=2\pi R$.

Let us first recapitulate some basic properties of the
one-dimensional diffusion with resetting. In order to fully describe
a resetting process by a Fokker-Planck equation formalism we need to
define different potentials describing the dynamics of the resetting
process. One needs to define a resetting potential
$\mathcal{P}_S(x)$ responsible for the particle removal associated
to the relocation of the particles. This potential is
in turn complemented by the gain potential $\mathcal{P}_G(x)$, which is
responsible for the reappearance of the particles at a different
position derived from the corresponding distribution function. Finally, as we are
dealing with a target process, we need to insert an annihilation
potential $\mathcal{P}_A(x)$.

This allows us to describe the evolution of the probability density
function $\psi(x,t;x_0)$ for a process starting from the position
$x_0$ at the time-point $t=0$, by solving the
master equation
\begin{widetext}
\begin{equation}
\frac{\partial \psi(x,t;x_0)}{\partial t} = D\frac{\partial^2
\psi(x,t;x_0)}{\partial x^2}+\mathcal{P}_G(x)\int \, \mathrm{d}x' \,
\mathcal{P}_S(x')\psi(x',t;x_0)-\mathcal{P}_S(x)\psi(x,t;x_0)
-\mathcal{P}_A(x)\psi(x,t;x_0).
\label{differentialone}
\end{equation}
\end{widetext}
This Fokker-Planck equation above can be generally used in order to
describe the dynamics for a large range of search processes. The
insertion of the potential $\mathcal{P}_A$ allows us to study even
processes with partial absorbing conditions \cite{Sano,Whitehouse,Scha}.
Since in the presented problems perfect absorbing boundary
conditions are implied, the annihilation potential $\mathcal{P}_A$
can be replaced by the boundary conditions
\begin{equation}
\psi(0,t;x_0)=\psi\left( L,t;x_0\right)=0.
\end{equation}
From the Fokker-Planck equation one can easily derive the backward
master equation fulfilled by the survival probability $Q(x_0,t)$,
\begin{equation}
Q(x_0,t)=\int \, \mathrm{d}x \, \psi(x,t;x_0)
\end{equation}
describing the probability for a process to have survived up to the
time-point $t$ if it started from the position $x_0$. It is indeed
possible to derive $Q(x_0,t)$ for a large range of potentials
describing complex dynamics, as we shall encounter in
Sec.~\textbf{IV}.

The first process which we deal with is the one-dimensional process
returning to its initial position after each reset with a uniform
resetting field and two perfect absorbing boundaries. In order to
solve this problem we are assuming now that $x_r$ is the position of
the particle after the reset. So we set
$\mathcal{P}_G(x)=\delta(x-x_r)$ and $\mathcal{P}_S(x)=r$ with
$\psi(0,x_0;t)=\psi(L,x_0;t)=0$.

It is not hard to see that in this case the backward master equation
(\ref{differentialone}) can be replaced by following expression
\begin{equation}
\frac{\partial Q(x,t)}{\partial t}= D_1\frac{\partial^2
Q(x,t)}{\partial x^2}-rQ(x,t)+rQ(x_0,t)
\label{survival}
\end{equation}
with
\begin{equation}
Q(0,t)=Q(L,t)=0.
\end{equation}
In order to solve this differential equation we can use the
Laplace-Transform
\begin{equation}
\tilde{q}(x,s)=\int\limits_{0}^{\infty} \, \mathrm{d}t \, Q(x,t)e^{-st}
\end{equation}
leading to
\begin{equation}
D_1\frac{\partial^2 \tilde{q}(x,s)}{\partial
x^2}-(r+s)\tilde{q}(x,s)+1+r\tilde{q}(x_r,s)=0.
\end{equation}
which yields \cite{Evans14}
\begin{equation}
\tilde{q}(x,s)=Ae^{\sqrt{\frac{r+s}{D_1}}x} +
Be^{-\sqrt{\frac{r+s}{D_1}}x}+\frac{1+r\tilde{q}(x_r,s)}{r}.
\end{equation}
Now by setting $s=0$ we are in the position to calculate the mean
time to absorption
\begin{equation}
T(x)=-\int_{0}^{\infty} \, \mathrm{d}t \,
t\frac{\partial Q(x,t)}{\partial t}=\tilde{q}(x,0)
\end{equation}
for which the expression
\begin{equation}
T(x)=Ae^{\ell^{-1} x}+Be^{-\ell^{-1} x}+\frac{1+rT(x_r)}{r}
\end{equation}
with $T(0)=T(L)=0$ holds. We used hereby the notation
\begin{equation}
\ell=\sqrt{\frac{D_1}{r}}
\label{mfpl}
\end{equation}
for the mean free path length between two resets.

We now can make the assumption that
\begin{equation}
\frac{\partial}{\partial x}T(x)|_{x=L/2}=0
\label{maxcon}
\end{equation}
which together with the two boundary conditions leads to the solution
\begin{equation}
T(x)=\frac{1+rT(x_r)}{r}\left(1-\frac{\cosh \ell^{-1}(x-L/2)}{\cosh \ell^{-1}L/2}\right)
\end{equation}
For the special case of $x_r=x$, namely the case where the particle
resets to its initial position, we can derive the formula
\begin{equation}
T(x)=-\frac{\ell^{2}}{D_1}\left(1-\frac{\sinh \ell^{-1}L}{\sinh \ell^{-1}x +\sinh \ell^{-1} (L-x)}\right).
\end{equation}
This result allows us to address the question with regard to the
dependence of the mean time on the starting position.

\begin{figure}
\input{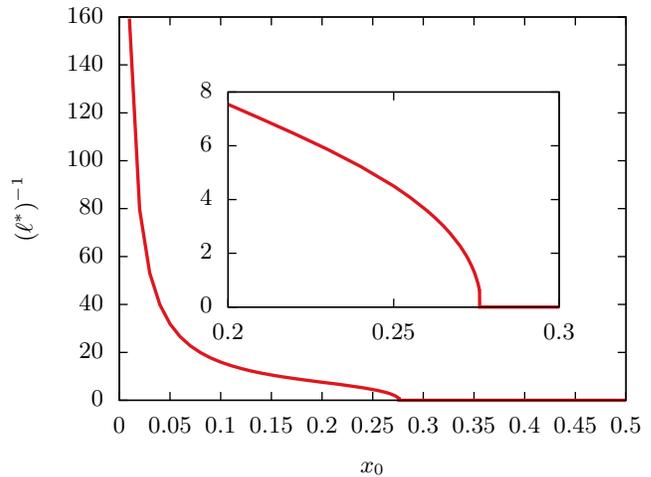}
\caption{Inverse of the optimal mean free path length vs the
starting  position of the searcher for a system of length $L=1$. We
can see that for $0.276L<x_0<0.724L$ the optimal mean free path
length is equal to $\infty$. This fact changes if we consider a
starting position which is closer to the boundaries and thus a
finite value for the mean free path length tends to be optimal.}
\label{swrdx0}
\end{figure}

We define therefore the optimal free path length, $\ell^{*}$, as
the value of $\ell$ for which
\begin{equation}
\frac{\partial}{\partial \ell}T(x)|_{\ell=\ell^{*}}=0
\end{equation}
holds. In Fig.~\ref{swrdx0} we can see that by decreasing the
distance between the starting position and one of the boundaries a
crossover for the optimal mean free path length from an infinite
value to a finite one takes place. This is particularly interesting from a physical point of view.
For initial positions near the boundaries, a finite traveling distance between two resets and hence $r>0$ is efficient. As soon as the particle starts at a position between the crossover points $(0.276L,0.724L)$, an infinite time interval between resets and thus pure diffusive motion is preferred. 

%%%%%%%%%%%%%%%%%%%%%%%%%%%%%%%%%%%%%%%%%%%%%%%%%%%%%%%%%%%%%%%%%%%%%%%%%%%%%%%%%

\section{Hard Resetting}

Let us now consider the case where the initial position is at the
center of the circle. We will characterize this process from now on
as {\it hard reset process}. It can be treated as a special case of
the general process presented in the next section. We treat this
special case here first since the presented results confirm previous
discoveries which were derived for processes that exhibit
intermittent dynamics \cite{Benichou10,Benicou14}. Furthermore
this first problem is actually rather simple and thus a good
introduction to our general approach.

Since the particle finds itself in the center of the circle after
each reset the subsequent one-dimensional excursion will start from
an arbitrary point on the circle due to the radial symmetry. This
allows us to use the uniform probability distribution function $\mathcal{P}_G(x)=L^{-1}=\left( 2\pi R \right)^{-1}$ $\forall x\in
[0,2\pi R]$ for the gain potential in equation
(\ref{differentialone}).

In order to calculate the MTA we will have to calculate the mean time for each of the two
different modes consisting purely of one- and two-dimensional
diffusion. Let $T_1$ be the mean time the particle
spends during its one-dimensional diffusion and $T_2$ the mean time
for the two-dimensional path then the total mean time to absorption
is given by,
\begin{equation}
T=T_1+T_2.
\end{equation}

For this simple process the MTA is given by
\begin{equation}
T=rT_1\tau_2+T_1+\tau_2, \label{generalmeantime}
\end{equation}
where $\tau_2$ is the mean time that a Brownian particle starting at
the center of a circle with radius $R$ needs in order to reach the boundary (mean time to boundary, MTB). This quantity is easily derived from the expression \cite{Redner01}
\begin{equation}
\tau_2(R)=\frac{R^2}{4D_2}.
\label{taur2d2}
\end{equation}

In order to determine the MTA of the
one-dimensional diffusion we just need to find the solution to the
differential equation
\begin{equation}
-1=D_1\frac{\partial^2}{\partial x^2}T_1(x)-\frac{r}{L}T_1(x)+\frac{r}{L}\int \, \mathrm{d}z \, T(z)
\end{equation}
with the boundary conditions
\begin{equation}
T(0)=T(L)=0.
\end{equation}
Which as shown in the previous section takes the form
\begin{equation}
T(x)=-A\frac{\cosh\ell^{-1}(x-L/2)}{r\cosh L
\left(2\ell\right)^{-1}}+\frac{L}{r}+F(0,L).
\label{GenT}
\end{equation}
where we used the notation $F(0,L)=\int_0^L \, \mathrm{d}z \, T(z)$.
From the boundary  conditions we can determine the constant $A$ as,
\begin{equation}
A=rF(0,L)+L.
\end{equation}
Reinserting the above expression in the general form described in (\ref{GenT})
and evaluating  $F(0,L)$ gives us
\begin{equation}
F(0,L)=\left(\frac{L}{r}+F(0,L)\right)\left(1-2\ell
\tanh L\left(2\ell\right)^{-1}\right)\nonumber
\end{equation}
from which we obtain
\begin{equation}
F(0,L)=\frac{L}{r}\left( \left(2\ell\right)^{-1} \coth
L\left(2\ell\right)^{-1}-1\right)\,.
\end{equation}
Finally we have
\begin{equation}
T(x)= \frac{L\coth L\left(2\ell\right)^{-1}}{2\ell r}\left(
1-\frac{\cosh\ell^{-1}(x-L/2)}{\cosh L\left(2\ell\right)^{-1}}\right).
\label{meantimeforurf}
\end{equation}

%%%%%%%%%%%%%%%%%%
\begin{figure}
\input{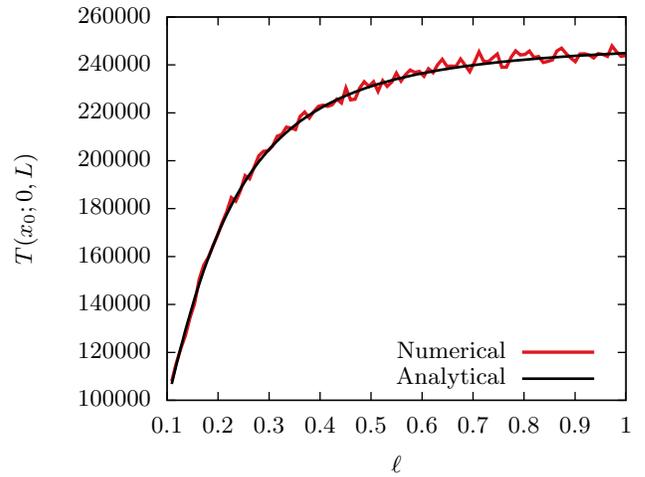}
\caption{Numerical evaluation of the formula (\ref{meantimeforurf})
for $x_0=0.5$ with $L=1$. We have chosen here, as before,
$D_1=5\cdot 10^{-7}$.}
\end{figure}
%%%%%%%%%%%%%%%%%%

From (\ref{meantimeforurf}) one can see that the mean time to
absorption is minimized for $\ell \rightarrow 0$. In order to prove
this result we use the following variation of the
original problem. We consider the gain potential
\begin{equation}
\mathcal{P}_G(x)=
\begin{cases}
    \frac{1}{1-2\varepsilon}      & \quad \text{if } \varepsilon<x<1-\varepsilon\\
    0  & \quad \text{else } \\
  \end{cases}
  \label{varepsilonpotential}
\end{equation}
which restores the original uniform potential in the limit
$\varepsilon\rightarrow 0$.  For the potential described by equation
(\ref{varepsilonpotential}) and $L=1$ we get the mean time
\begin{equation}
T(x)= \frac{(1-2\varepsilon)\left( \cosh\left(2\ell\right)^{-1}
-\cosh\ell^{-1}(x-1/2)\right)}{2r\ell\sinh\ell^{-1}(1/2-\varepsilon)}.
\label{Tvarepsi}
\end{equation}
For small values of $\ell$ and $x\in[0,1]$ we
have
\begin{equation}
\left( 1-\frac{\cosh\ell^{-1}(x-1/2)}{\cosh\left(2\ell\right)^{-1}}\right)\simeq 1.
\end{equation}
This allows us to replace equation (\ref{Tvarepsi}) above with the expression
\begin{equation}
T(x)= \frac{\sigma^{-2}2(1-2\varepsilon)\ell}{\tanh\left(2\ell\right)^{-1}-
\cosh \ell^{-1}\varepsilon-\sinh \ell^{-1}\varepsilon}.\nonumber
\end{equation}
Now by using
\begin{equation}
\tanh\left(\frac{\ell^{-1}}{2}\right)\simeq 1, 
\end{equation}
for small values of $\varepsilon$ we can simplify our formula to
\begin{equation}
T(x)= \varepsilon\sigma^{-2}(1-2\varepsilon)
\frac{e^{\ell^{-1}\varepsilon}}{\ell^{-1}\varepsilon}.
\end{equation}
We also see that for high values of the resetting rate, the mean
time becomes independent of the initial position since due to the
resetting, the information with regard to the initial position gets
lost for increasing times.

Now we need to consider that the function $x^{-1}e^{x}$ has a
minimum for $x=1$.  Correspondingly, the optimal resetting rate for
this one-dimensional problem is achieved when
$\ell^{-1}=\varepsilon^{-1}$. By sending now $\varepsilon$ to zero,
in order to regain the original problem, we notice that the optimal
resetting rate goes to infinity.

\begin{figure}
\input{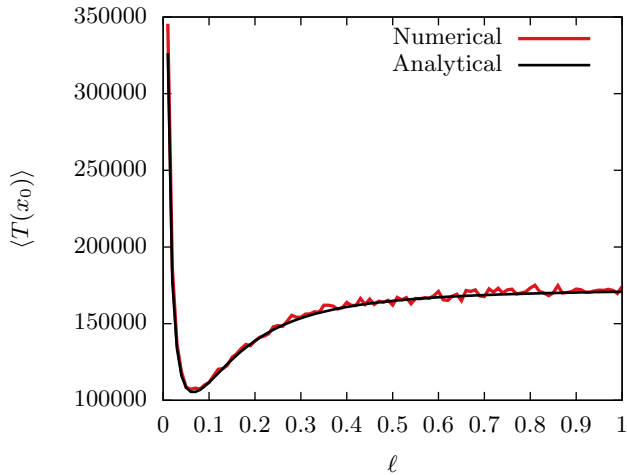}
\caption{MTA for the hard resetting problem in the special
case  of $D_2/ D_1=2$ for $R=(2\pi)^{-1}$ and $D_1=5\cdot 10^{-7}$.}
\label{D2D12}
\end{figure}

\begin{figure}
\input{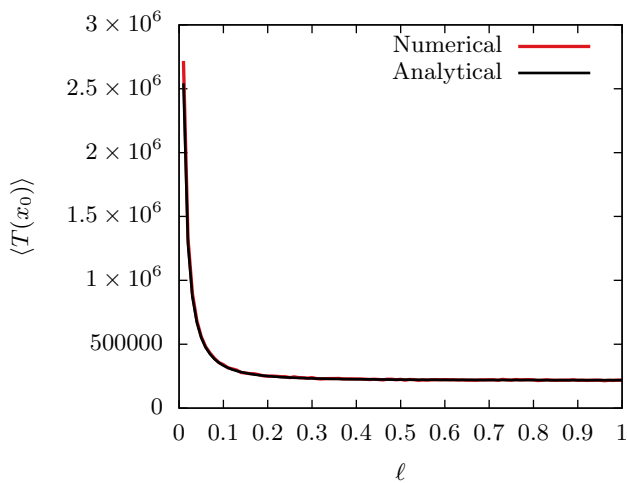}
\caption{MTA for the hard resetting problem in the special
case of  $D_2/ D_1=0.25$ for $R=(2\pi)^{-1}$ and $D_1=5\cdot
10^{-7}$.} \label{D2D1025}
\end{figure}

Although there is no finite optimal resetting rate for the
one-dimensional case, the situation changes when we consider the
two-dimensional excursion of the particle. We evaluate therefore
Eq.~(\ref{generalmeantime}) by implementing Eq.~(\ref{taur2d2}) and
Eq.~(\ref{meantimeforurf}) and integrating over the interval $[0,L]$
leading to
\begin{equation}
T=\frac{R^2}{4D_2}+\frac{\left(L\coth L\left(2\ell \right)^{-1}-2\ell\right)
}{2r\ell}\left(\frac{rR^2}{4D_2}+1\right).
\label{D2D1}
\end{equation}

This expression is in perfect agreement with Eq.~(2.23) of
\cite{Benicou14}. By using the series expansion of the cotangent
function we rewrite this expression as
\begin{equation}
T=\frac{R^2}{4D_2}+\frac{2}{D_1}\left(1+\frac{r}{D_2}\right)
\sum\limits_{k=0}^{\infty} \frac{L^{-1}}{\left(\frac{2\pi k}{L}\right)^2+\frac{r}{D_1}}.
\end{equation}
It follows from this result that an optimal resetting rate can be
found only if $D_2/D_1\geq 0.38$. This result is easily obtained by a numerical evaluation of the formula above. In Figs.~\ref{D2D12} and \ref{D2D1025} we can see the
diverse behavior of the mean time to absorption for different values
of $\ell$ for values above and below the critical value of
$D_2=0.38D_1$ respectively.

%%%%%%%%%%%%%%%%%%%%%%%%%%%%%%%%%%%%%%%%%%%%%%%%%%%%%%%%%%%%%%%%%%%%%%%%%%%%%%%%%

\section{Partial Resetting}

Now we consider the general case where the initial position lies
inside the circle and is not the center. We start with the problem
where resetting can only take place from the boundary of the domain. We will characterize this strategy from now on as {\it partial reset process}. In the next section we will consider the problem of a persistent resetting field, where resetting can take place from anywhere during the Brownian motion 
of the particle, including from aywhere inside the circle, irrespective of whether the particle
is undergoing one-dimensional or two-dimensional diffusion.

Let us first imagine that our particle starts from the position, given by the
polar coordinates $\vec{x}=\left(R_0,\theta_0\right)$, where $R_0$ is the distance from the center
and $\theta_0$ is the angle it makes with the vertical. After a two-dimensional excursion the particle arrives at the circle.
The mean time for this arrival at the boundary is given by
\begin{equation}
\tau_2(R,R_0)=\frac{R^2-R_0^2}{4D_2}.
\end{equation}
Depending on the parameters $R_0$ and $\theta_0$ the likelihood of
crossing the boundary at a specific angle has to be calculated. This
allows us to determine the probability distribution function of the
starting position for the subsequent one-dimensional diffusion
phase. In order to compute the probability distribution function of
the hitting angle we can use an electrostatics analogy
\cite{Redner01} which leads to
\begin{equation}
 \mathcal{P}(\theta;\theta_0,R_0) = \frac{1-\frac{R_0^2}{R^2}}{2\pi R
 \Big(1-\frac{2R_0\cos(\theta-\theta_0)}{R} + \frac{R_0^2}{R^2}\Big)}.
 \label{probabilitytheta}
\end{equation}

%%%%%%%%%%%%%%%
\begin{figure}
\input{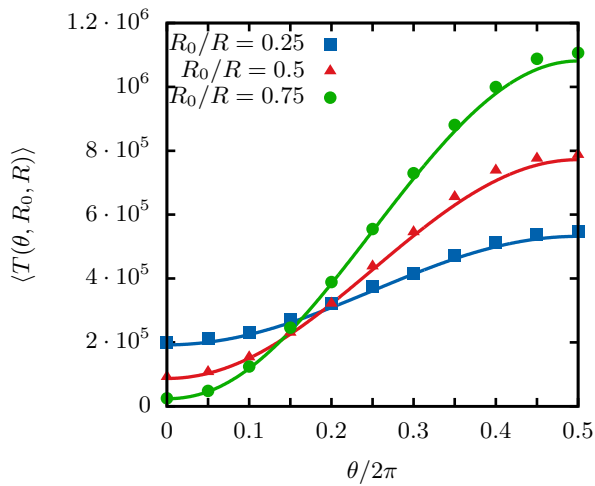}
\caption{Mean time to absorption for a Brownian process starting
inside a two-dimensional circle for different initial conditions
$(R_0,\theta)$. The points were determined by a Monte-Carlo
simulation with $D_2=10^{-6}$, $2D_1=D_2$ and $R=1/2\pi$. The lines
correspond to our theoretical values for these parameters.}
\label{D2D1rT}
\end{figure}
%%%%%%%%%%%%%%

Let $x_0$ be the point on the boundary of the circle that lies on
the line connecting  the initial position
$\vec{R}_0=\left(\theta_0,R_0\right)$ to the center of the circle
and is closest to the initial position. Then
$\mathcal{P}_G(z)=\mathcal{P}(\theta,R_0)$ holds, if we assume that
$R (\theta-\theta_0)=(z-x_0)$. Inserting this expression in
Eq.~(\ref{differentialone}) we can get following backward equation
for the mean time to absorption for a process starting on the
boundary at the position $x$,
\begin{equation}
0=D_1\frac{\partial^2}{\partial x^2}T(x)-rT(x)+1+r\int \, \mathrm{d}z \, P_G(z)T(z).
\end{equation}
Using the boundary condition $T(0)=T(2\pi R)=0$ together with
Eq.~(\ref{maxcon}) we can easily solve this non-homogeneous
differential equation of second order to deliver
\begin{equation}
T(x)=\left(\frac{1}{r}+\int \, \mathrm{d}z \,
\mathcal{P}_G(z)T(z)\right) \left(1-\frac{\cosh\ell^{-1}(x-\pi
R)}{\cosh \ell^{-1}\pi R}\right).
\end{equation}
If we multiply now both sides of this formula with
$\mathcal{P}_G(x)$ and integrate  over the interval $[0,2\pi R]$ we
can see that
\begin{equation}
T(x)=\frac{\cosh \sqrt{\frac{r}{D}}\frac{L}{2}-\cosh \sqrt{\frac{r}{D}}
\left(x-L/2\right)}{r\int \, \mathrm{d}z \, \mathcal{P}_G(z)\cosh
\sqrt{\frac{r}{D}}\left(z-L/2\right)}
\end{equation}
holds.

%%%%%%%%%%%%%%
\begin{figure}
\includegraphics[width=0.5\textwidth]{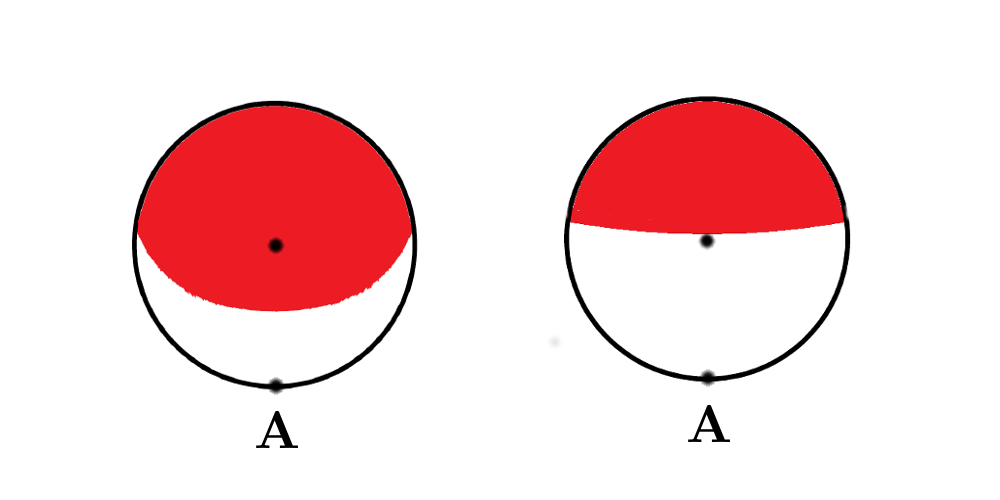}
\caption{Different regimes with regard to the existence of an
optimal resetting rate for different initial position for a Brownian particle in search for a target located at \textbf{A}. The left figure corresponds to the choice of parameters with $D_2/D_1=0.25$ and on the right to $D_2=D_1$. The shaded area (red) describes initial positions for which no positive optimal resetting rate can be found. We can see that for $D_2/D_1=1$ a particle starting from the center of the circle can be optimized through a positive resetting rate.} \label{figfig}
\end{figure}
%%%%%%%%%%%%

The MTA is thus given by the expression
\begin{equation}
T_1 =\int \, \mathrm{d}z \, \mathcal{P}_G(z)T(z).
\end{equation}
Implementing all of these results in Eq. (\ref{generalmeantime}) we get
\begin{equation}
T=\frac{R^2-R_0^2}{4D_2}+\left(\frac{1}{r}+\frac{R^2-R_0^2}{4D_2}\right)\times\nonumber
\end{equation}
\begin{equation}
\times\int \, \mathrm{d}x \, \mathcal{P}_G(x)\frac{\cosh \sqrt{\frac{r}{D_1}}\pi
R-\cosh \sqrt{\frac{r}{D_1}}\left(x-\pi R\right)}{\int \, \mathrm{d}z \,
\mathcal{P}_G(z)\cosh \sqrt{\frac{r}{D_1}}\left(z-\pi R\right)}.
\label{sectioniv}
\end{equation}
In Fig.~\ref{D2D1rT} we can observe a very good agreement between
this formula and the results of Monte-Carlo simulations.

The derived equation gives us the opportunity to characterize the
different regimes inside the circle for which an optimal resetting
rate can be found. In Fig.~\ref{figfig} we represent the respective
areas for the ratios of $D_2=0.25D_1$ and $D_2=D_1$ which are below
and above the characteristic value of $D_2=0.38D_1$, that was
determined in the last section. In both figures the $R_0/R=1$ axis
is crossed at the value of $\theta=0.56\pi$, which is in perfect
agreement with the results of Sec.~\textbf{II}.

%%%%%%%%%%%%%%%%%%%%%%%%%%%%%%%%%%%%%%%%%%%%%%%%%%%%%%%%%%%%%%%%%%%%%%%%%%%%%%%%%

\section{Persistent Resetting}

In the previous section we could determine for specific values of the ratio $D_2/D_1$ the different regimes for
which a resetting from the boundary could be beneficial.
In this section we analyze the dynamics for a process
for which resetting takes place not only from the boundary, but also
from inside of the circle. We consider also a permanent resetting field and characterize in the following the process as {\it persistent reset process}.

While the formalism and approach of the previous sections proves
useful, special care  with regard to two characteristics of the
present problem have to be taken into account. The first is the
modification of the MTB, $\tau_2$, due to the resetting mechanism.
The second is the effect of the resetting mechanism on the hitting
angle distribution.

We start with the first point, the goal being to determine the
conditions under which an optimal resetting rate with regard to
$\tau_2$ can be found. Let $\vec{x}_r$ be the position of resetting
then the two-dimensional excursions of the particle are described by
the backward equation
\begin{equation}
D_2\nabla^2 \tilde{q}(\vec{x}, s)-(r+s)\tilde{q}(\vec{x}, s)=-1-r\tilde{q}( \vec{x}_r, s)
\label{2Ddiffeq}
\end{equation}
with the boundary condition
\begin{equation}
\tilde{q}\left( \vec{x}, s \right)|_{|\vec{x}|=R}=0.
\end{equation}

We start by providing the solution to the homogeneous equation
\begin{equation}
D_2\nabla^2 \tilde{q}(\vec{x}, s)-(r+s)\tilde{q}(\vec{x}, s)=0
\end{equation}
which is radially symmetric about the origin and has a vanishing
derivative at $|\vec{x}|=0$
\begin{equation}
\nabla \tilde{q}(\vec{x}, s)|_{\vec{x}=\vec{0}}=0.
\end{equation}
These conditions are fulfilled by the equation
\begin{equation}
\tilde{q}_{hom}(\vec{x}, s)=I_0\left( \sqrt{\frac{r+s}{D_2}}|\vec{x}|\right)
\end{equation}
where $I_n$ is the modified Bessel function of the first type.

Now we can use this expression in order to solve the differential
equation (\ref{2Ddiffeq}), by considering, as before, the Ansatz
\begin{equation}
\tilde{q}\left( \vec{x}, s \right)=AI_0\left( \sqrt{\frac{r+s}{D_2}}|\vec{x}|\right)+B.
\end{equation}
By taking into account the boundary condition we can derive the solution
\begin{equation}
\tilde{q}\left( \vec{x},
s\right)=\frac{I_0\left(\ell^{-1}_sR\right)-I_0
\left(\ell^{-1}_s|\vec{x}|\right)}{rI_0\left(\ell^{-1}_s|\vec{x}_r|\right)
+sI_0\left(\ell^{-1}_sR\right)}
\end{equation}
with $\ell^{-1}_s=\sqrt{(r+s)/D_2}$. For the problems studied in
this paper, where the  resetting position $\vec{x}_r$ is equal to
the starting position, we have
\begin{equation}
\tilde{q}\left( \vec{x},
s\right)=\frac{I_0\left(\ell^{-1}_sR\right)-
I_0\left(\ell^{-1}_s|\vec{x}|\right)}{rI_0\left(\ell^{-1}_s|\vec{x}|\right)
+sI_0\left(\ell^{-1}_sR\right)}.
\end{equation}

\begin{figure}
\input{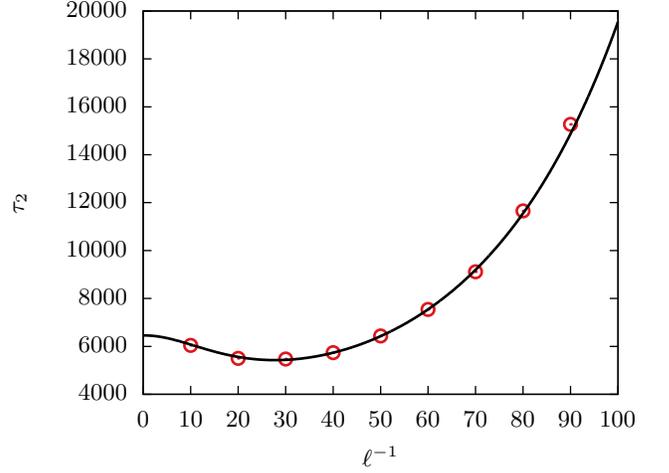}
\caption{MTB for two-dimensional
Brownian motion in a  circle of radius $R=1$ starting from the
radius $R_0=0.9$ for different inverse mean path lengths. We can
clearly see that there exists an optimal resetting rate for which
this time is minimized.} \label{D2D1R0R07}
\end{figure}

\begin{figure}
\input{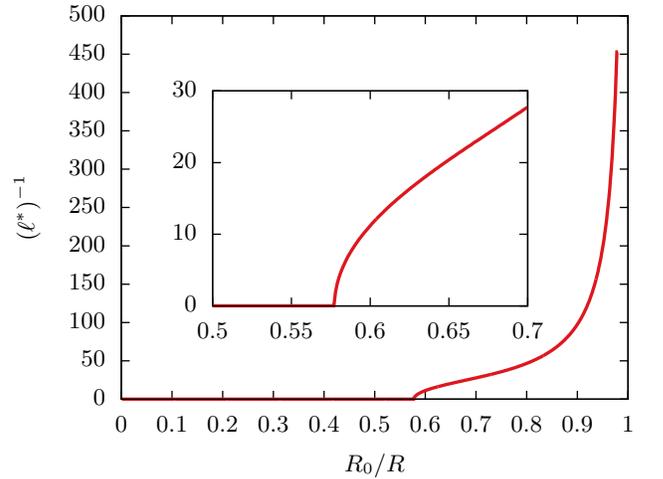}
\caption{Inverse of the optimal mean free path length for different
ratios of the  initial radius to the total radius of the circle. We
can see that for $R_0>0.578R$ the optimal resetting rate is greater
than zero. We have used here a circle of radius $R=(2\pi)^{-1}$.}
\label{Besselnew6}
\end{figure}

Now we can determine the MTB for this process
provided  by the formula
\begin{equation}
\tau_2(R_0)=\tilde{q}\left( R_0,
s=0\right)=\frac{1}{r}\left(\frac{I_0
\left(\ell^{-1}_2R\right)}{I_0\left(\ell^{-1}_2R_0\right)}-1\right)
\label{tau2(R0)}
\end{equation}
with $\ell_2=\sqrt{D_2/r}$. In Fig,~\ref{D2D1R0R07} we can see a
very good agreement  between the derived analytical solution and the
Monte-Carlo simulation. Furthermore in Fig.~\ref{Besselnew6} we
determine the optimal resetting rate for a large range of values of
the ratio $R_0/R$. As we can see, an optimal resetting rate can be
found only for $R_0>0.578R$.

We can even calculate the ratio $D_2/D_1$ for which the effect of
resetting is beneficial when starting from the center of the circle.
We can use the same formula as before to show that the mean time to
absorption in this case is given by
\begin{equation}
T=\frac{1}{r}\left( \frac{I_0\left(\ell^{-1}_2R\right)}{I_0\left(\ell^{-1}_2R_0\right)}
-1\right)+\frac{I_0\left(\ell^{-1}_2R\right)}{I_0\left(\ell^{-1}_2R_0\right)}\times\nonumber
\end{equation}
\begin{equation}
\times\frac{\coth L(2\ell_1)^{-1}}{2r\ell_1}\left(L-2\ell_1\tanh L\left(2\ell_1\right)^{-1}\right)
\label{finaleq2}
\end{equation}
with $\ell_1=\sqrt{D_1/r}$. By evaluating this formula we can see
that a necessary condition  for the existence of an optimal
resetting rate is $D_2>0.4D_1$. As expected, this value is
slightly larger than the value ($D_2>0.38D_1$) calculated in the
last section for the special case where the resetting field acts
only on the boundary of the system.

Now we focus on the effect of the resetting on the hitting angle. As
discussed in  \cite{Belan} a resetting mechanism has a significant
effect on the outcome of an absorption process. This can be
understood by considering the stochastic paths of the particle
influenced by resetting as the paths of a renewal process. The mean
length of these paths is proportional to $r^{-1}$. Successfully
approaching the boundary is only possible for the direct paths
between initial position and boundary.

We use the method introduced in \cite{Belan} in order to calculate
the effect.  Let $P(\theta,t)$ describe the probability density
function of the absorption time for a path that crosses the boundary
at the point corresponding to the angle $\theta$, then the probability for the same outcome
with a positive resetting rate $r$ is given by
\begin{equation}
P_r(\theta)=\frac{\tilde{P}(\theta,r)}{\tilde{P}(r)},
\end{equation}
where $\tilde{P}(\theta,r)$ and $\tilde{P}(r)$ denote the Laplace
Transforms of   $P(\theta,t)$ and $P(t)=\int \, \mathrm{d}\theta \,
P(\theta,t)$ evaluated at $r$, respectively. In order to calculate
the probability distribution functions $P(\theta,t)$, the image
method proves cumbersome. We therefore suggest a Monte-Carlo simulation for the calculation of this probability distribution function.

The calculation of the Laplace Transform $\tilde{P}(r)$ is easier on
the  other side, since
\begin{equation}
\tilde{P}(r)=-\int\limits_{0}^{\infty} \, \mathrm{d}t \,
\frac{\partial Q(R_0,t)}{\partial t}e^{-rt}=\nonumber
\end{equation}
\begin{equation}
=-e^{-rt}Q(R_0,t)|_0^{\infty}-r\int\limits_{0}^{\infty} \, \mathrm{d}t \,
e^{-rt}Q(R_0,t)=\nonumber
\end{equation}
\begin{equation}
=\frac{2I_0(\ell^{-1}_2R_0)}{I_0(\ell^{-1}R)+I_0(\ell^{-1}_2R_0)}.
\end{equation}

As in Sec.~\textbf{IV} we can use this formula in order to determine
the probability distribution function of the starting point on the
circle $\mathcal{P}_G(z)=P_r(\theta)$ by setting
$R(\theta-\theta_0)=(z-x_0)$.

This way we can summarize all results of the present section in the
following formula
\begin{equation}
T=\frac{1}{r}\left(
\frac{I_0\left(\ell^{-1}_2R\right)}{I_0\left(\ell^{-1}_2R_0\right)}
-1\right)+\frac{I_0\left(\ell^{-1}_2R\right)}{I_0\left(\ell^{-1}_2R_0\right)}\times\nonumber
\end{equation}
\begin{equation}
\times\int \, \mathrm{d}x \, \mathcal{P}_G(x)\frac{\cosh
\ell^{-1}_1\pi R- \cosh \ell^{-1}_1\left(x-\pi R\right)}{\int \,
\mathrm{d}z \, \mathcal{P}_G(z)\cosh \ell^{-1}_1\left(z-\pi
R\right)}.
\label{finaleq}
\end{equation}
This expression is in good agreement with our previous results. For
example it is not hard  to see that in the limit
$D_2\rightarrow\infty$ the present formula is equivalent to
Eq.~(\ref{sectioniv}).

\begin{figure}
\input{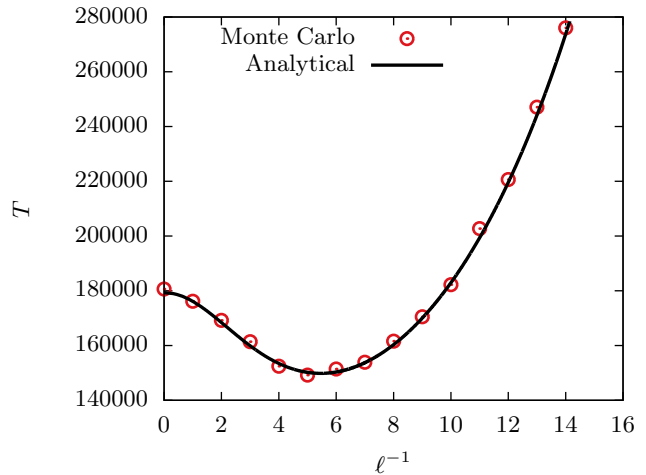}
\caption{MTA for a particle starting from the center of the circle of radius $R=(2\pi)^{-1}$ for different inverse mean path lengths. We can
clearly see that there exists an optimal resetting rate for which
this time is minimized. The results of the Monte-Carlo simulation show a very good agreement to the analytical expectations derived from the formula (\ref{finaleq2}).}
\label{lastfig}
\end{figure}

%%%%%%%%%%%%%%%%%%%%%%%%%%%%%%%%%%%%%%%%%%%%%%%%%%%%%%%%%%%%%%%%%%%%%%%%%%%%%%%%%

\section{conclusion}

In this work we studied the dynamics of a diffusive searcher which
combines two different behaviors. In one mode the searcher undergoes
a two-dimensional diffusion and in the second mode a one-dimensional
diffusion process along the boundary is performed. We also assumed
that the evolution of the process is accompanied by restarts
allowing the searcher to return to its initial position.

We started by a Fokker-Planck formalism and used the Laplace
Transform to calculate the mean time to absorption for several
variations of the process. This allowed us to get a better
understanding of the properties of random search processes with
restarts in two-dimensional bounded domains. In the past we could
show that resetting can be beneficial for the case of an
one-dimensional diffusion process in a bounded domain \cite{Scha}.
The work presented here allows us to claim that this statement holds
also for a diffusion process inside a two-dimensional domain.

Specifically we could see that for the case of a one-dimensional
process taking place in a periodic domain of length $L$ an optimal
resetting rate exists as long as the distance between the initial
position and the target is smaller than $0.276L$. In the next
section we found that if the resetting distribution applies
on the whole interval $[0,L]$, then the optimal resetting rate is
equal to $\infty$.

These results were especially useful when analyzing the process
switching from two-dimensional diffusion to one-dimensional
excursion on the boundary where a resetting field applies. For this
problem we could determine several regions for which an optimal
resetting rate that was bigger than zero for different values of the
ratio $D_2/D_1$ between the two diffusion constants. It was also
possible to show that when the process starts from the center of the
circle an optimal resetting rate can be found as long as
$D_2>0.38D_1$. This is in perfect agreement with the past findings of
similar works concerned with processes exhibiting intermittent
behavior \cite{Benichou10}. In the last section we considered the
possibility of a reset from also inside the circle and have shown
that when starting from the center of the circle a resetting rate
that is bigger than zero can only be preferential when
$D_2>0.4D_1$.

Another generalization of the present formalism could consist in the
replacement of the point-like target by an extended area on the
boundary. The presented methods can easily be implemented to solve
the problem of a Brownian particle trying to escape from a bounded
domain through a small window while under the effect of a resetting
potential. This can be considered as a variation of the narrow
escape problem \cite{Voituriez}. It would be interesting to see if a
positive resetting rate can accelerate this process.

In several works in the past discussing diffusion with resetting,
the case of many searchers has also been
considered \cite{Evans11, Evans14, Whitehouse}. We think that such a
consideration would also be interesting in the present setting. One
could go even further and implement an interaction mechanism between
these searchers by adopting a single-file diffusion process on the
boundary of the circle.

%%%%%%%%%%%%%%%%%%%%%%%%%%%%%%%%%%%%%%%%%%%%%%%%%%%%%%%%%%%%%%%%%%%%%%%%%%%%%%%%%

\appendix*

\section{Monte-Carlo simulation}

\begin{figure}
\input{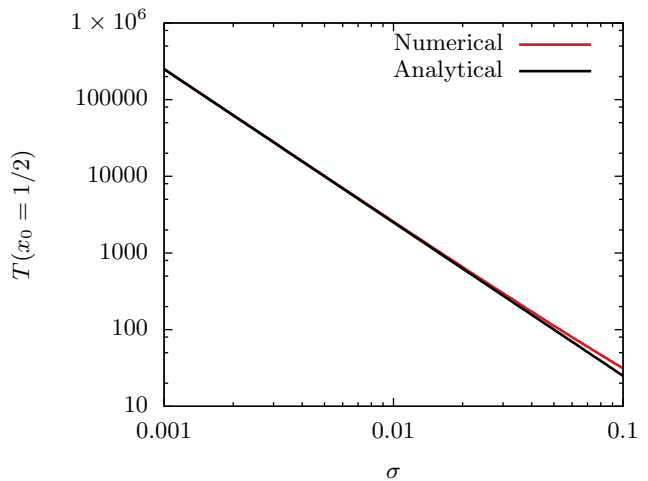}
\caption{Numerical and analytical evaluation of the targeting
problem  with no resetting. We calculate the mean time to absorption
for a Brownian particle starting from $x_0=1/2$ with two absorbing
boundary conditions $T(0)=T(1)=0$. We can see that a comparison
between our analytical and numerical results is only possible for
low values of $\sigma$. } \label{RWPB}
\end{figure}

In order to compare our analytical findings with numerical results
we  relied upon Monte-Carlo simulations. An easy way to simulate a
Brownian-like motion consists hereby of observing the evolution of a
discrete time random walk $x_t$ with $t\in\mathbb{N}$, defined by
\begin{equation}
x_t=x_0+\sum_{k=0}^t \xi_k
\end{equation}
whereas the random variables $\xi_k$ are drawn from a normal
distribution  with mean $0$ and variance $\sigma^2$.

The mean variance of the random walk is given by the formula
\begin{equation}
\mathbb{E}\left[x_t^2-x_0^2\right]=\mathbb{E}
\left[\sum\limits_{i=1}^t(x_i-x_{i-1})^2\right]=\sigma^2t.
\end{equation}
By comparing this expression to the Green's function we see that
$\sigma^2=2D_1$.  This relation between these two scales (analytical
and numerical) allows us to directly compare our numerical results
to the derived analytical expressions.

We have to note here that one has to be careful with regard to the
choice of  standard deviation, since higher values of volatility may
lead to discrepancies between the two approaches since the nature of
the continuum (analytical) and the discrete (numerical) time-step
processes is fundamentally different. This is shown in Figure
\ref{RWPB} where the expected time to absorption for a process with
no resetting and two perfectly absorbing boundary conditions is
plotted for both numerical and analytical methods.

In the latter sections \textbf{IV} and \textbf{V} the implementation
and  understanding of the dynamics of two-dimensional Brownian
motions became necessary. In order to achieve that, we considered a
two-dimensional random walk described by the equation
\begin{equation}
\vec{x}_t=\left(x_0+\sum_{k=0}^t\xi_k\right)\vec{e_x}+\left(y_0+\sum_{k=0}^t\xi_k\right)\vec{e_y},
\end{equation}
whereas $\xi_k$ is a white noise process derived from the Gaussian
distribution $\mathcal{N}(0,\sigma)$. For the presented
two-dimensional process we have a standard deviation given by
$2D_2=\sigma^2$. This allows us to determine numerically the
stopping time
\begin{equation}
\tau_2=\inf \left\{ t\in\mathbb{N}: x_t^2+y_t^2\geq R^2\right\}.
\end{equation}
For two-dimensional systems the offset is
extremely hard to reduce in comparison to the one-dimensional
processes. Fortunately we can rely hereby on two methods: using
processes with a smaller volatility and/or introduce a finer
time-scale. The first method is non-optimal since it leads to an
increase of the mean time to absorption and correspondingly to
higher running times.

We decided therefore to use the second method. We considered hereby
that each time step consists of  $\delta^{-2}$ smaller steps during
which jumps of length $\delta \xi$ are performed. It is easy to see
that by sending $\delta$ to zero the desired Wiener process can be
approximated, but even for $\delta>0$ this method leads to a great
improvement of the derived results. For section \textbf{IV} we have
chosen $\delta=0.25$ and for section \textbf{V} $\delta=0.1$.

In order to simulate the resetting mechanism we introduced the
resetting  probability $r_n\in[0,1]$. After each jump the walker is
reset to a new position with a probability $r_n$ according to the
gain distribution $\mathcal{P}_G$. The mean time between two resets
$\tau_r$ for this process is consequently given by the formula
\begin{equation}
\tau_r=1+\sum_{k=0}^{\infty}kr_n(1-r_n)^k=1+\frac{1-r_n}{r_n}=\frac{1}{r_n}.
\end{equation}
In our analytical calculations we know that the times intervals
between  two resets have an exponential distribution
\begin{equation}
P(t)=re^{-rt}
\end{equation}
for which then the mean time is given by $\tau_r=\frac{1}{r}$.

A comparison between our analytical findings and the Monte-Carlo
approach is made possible by adjusting the two parameters, $r_n$ and
$r$, so that these two time scales are equal.

\acknowledgements

One of the authors (AC), acknowledges the financial support received from Deutscher Akademischer Austauschdienst (DAAD) during the work. AC is also grateful to the University of Cologne for their kind hospitality. The work of CC and AS was supported by Deutsche Forschungsgemeinschaft under grant SCHA 638/8-2.

%%%%%%%%%%%%%%%%%%%%%%%%%%%%%%%%%%%%%%%%%%%%%%%%%%%%%%%%%%%%%%%%%%%%%%%%%%%%%%%%%%%%

\end{document}